\documentclass[11pt,twoside]{article}
\usepackage{asp2010}
\usepackage{ogonek}
\resetcounters

\begin{document}

\title{Measuring stellar rotation periods with Kepler}
\author{M. B. Nielsen$^1$$^,$$^2$, L. Gizon$^2$$^,$$^1$, H. Schunker$^2$, and C. Karoff$^3$
\affil{$^1$Institut f{\"u}r Astrophysik, Georg-August-Universit{\"a}t G{\"o}ttingen, Friedrich-Hund-Platz 1, 37077 G{\"o}ttingen, Germany}
\affil{$^2$Max-Planck-Institut f{\"u}r Sonnensystemforschung, Max-Planck-Stra{\ss}e 2, 37191 Katlenburg-Lindau, Germany}
\affil{$^3$Stellar Astrophysics Centre (SAC), Department of Physics and Astronomy, Aarhus University, 
             Ny Munkegade 120, DK-8000 Aarhus C, Denmark}}

\begin{abstract}
We measure rotation periods for 12\,151 stars in the {\it Kepler} field, based on the photometric variability caused by stellar activity. Our analysis returns stable rotation periods over at least six out of eight quarters of {\it Kepler} data. This large sample of stars enables us to study the rotation periods as a function of spectral type. We find good agreement with previous studies and $v \sin i$ measurements for F, G and K stars. Combining rotation periods, B-V color, and gyrochronology relations, we find that the cool stars in our sample are predominantly younger than $\sim1$\,Gyr.
\end{abstract}

\section{Introduction}
Stellar rotation is a key aspect of stellar evolution, stellar activity, and mass loss through stellar winds. The spin down experienced by stars during their main sequence lifetimes is also a measure of their ages \citep[e.g.][]{Barnes2007}. The \emph{Kepler} mission provides us with a unique opportunity to study stellar rotation over a large and consistent sample of stars covering a wide range of spectral types.

Among the high-quality photometric light curves of {\it Kepler}, a significant fraction of stars show periodicities associated with active regions. Using spectral analysis, we extract stable rotation periods from 8 quarters of {\it Kepler} observations. Here we summarize the results published by \citet{Nielsen2013} and further comment on the prospects for inferring ages of stars through gyrochronology.

\section{Method}
We use the white-light photometric {\it Kepler} time series, which consist of $\sim90$\,day segments (quarters). We analyzed quarters 2 to 9 for a total of $192\,668$ stars. We considered the data processed by PDC\_MAP reduction pipeline \citep{Smith2012} and also the more recent data from the msMAP pipeline \citep{Thompson2013} for comparison purposes. Periods longer than $\sim 21$\,days are treated differently by the two pipelines and so may lead to different stellar rotation periods \citep[Garcia private communication,][]{Nielsen2013}. 

For each star and each quarter we computed the Lomb-Scargle periodogram using PDC\_MAP data. We then identified the peak of maximum power in the period range 1\,-\,100 days, discarding periods longer than 30 days (which are potentially subject to instrumental effects). Detected periods are subjected to a statistically analysis in order to only extract stable periods, namely at least 6 periods must fall within 2 median absolute deviations (MAD), with a maximum MAD of 1 day. These stable periods are adopted as stellar rotation periods.

\section{Rotation Periods}
The PDC\_MAP rotation periods for 12\,151 stars are published as online material via the CDS \footnote{http://cdsarc.u-strasbg.fr/viz-bin/qcat?J/A+A/557/L10}. The msMAP data yielded only 9617 stars with measured rotation periods. The msMAP tends to suppress long term variability and is therefore ill suited for studying slow rotators. 

\begin{figure}[!t]
\centering
\includegraphics[width=0.8\columnwidth]{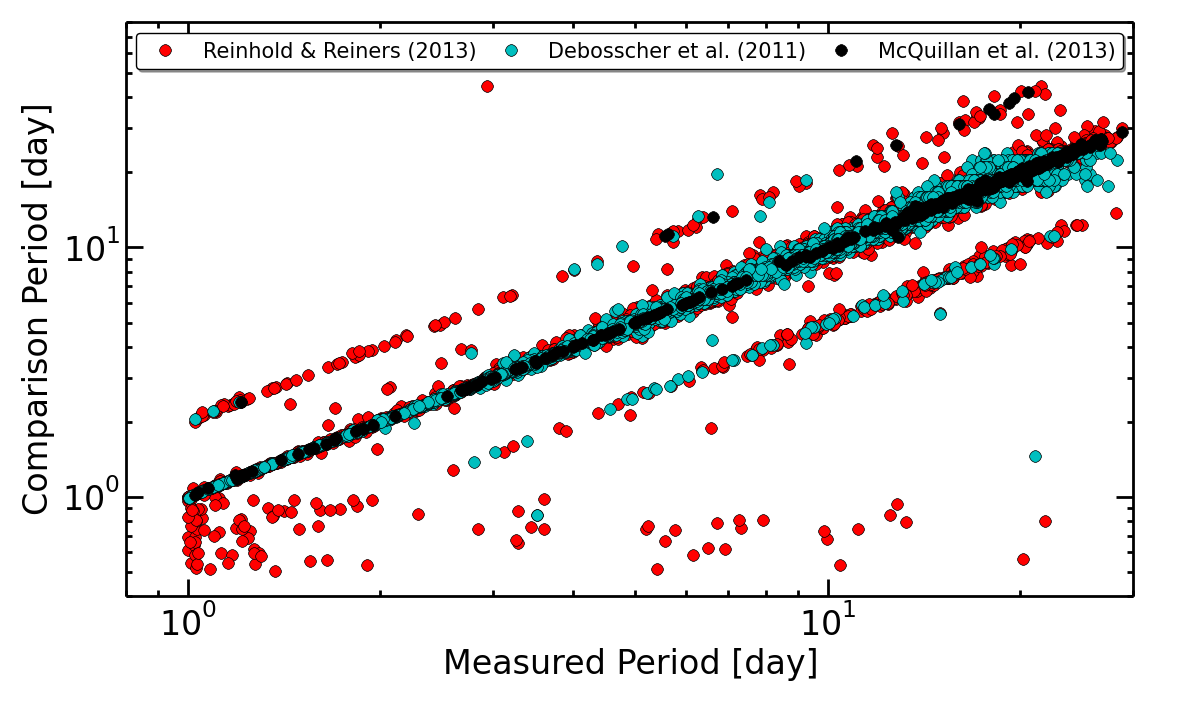}%
\caption{Comparison between the measured PDC\_MAP rotation periods (this work, abscissa) and periods from \citet{Reinhold2013} (red), \citet{Debosscher2011} (blue), and \citet{McQuillan2013} (black).}%
\label{fig:comparison}%
\end{figure}

We compared our results with those of \citet{Debosscher2011} (8654 stars), \citet{Reinhold2013} (24\,124 stars), and \citet{McQuillan2013} (1570 stars). In Figure~\ref{fig:comparison} we show periods for the stars in our sample that are also present in each of the respective comparison samples. It is clear that the vast majority ($>90\%$) of detected periods are consistent among the different samples. It is also evident that some periods differ, as seen by the points that lie at twice or half our measured period. \citet{Reinhold2013} and \citet{McQuillan2013} consider periods longer than our limit of $30$\,days, this means that a few of our periods above 30 days are likely harmonic periods. \citet{Debosscher2011} use a similar period range as ours, and so it is unclear why in some cases their analysis yields what appears to be sub-harmonic to our measured periods. Similarly \citet{Reinhold2013} analyzed periods shorter than one day, giving rise to the scatter of points in the lower part of the figure. The comparison periods from \citet{Reinhold2013} and \citet{Debosscher2011} that are found at half our measured period are likely caused be the fact that these studies only considered one quarter of {\it Kepler} data.

\section{Rotation vs. spectral type}
\citet{Nielsen2013} presented an additional comparison with $v \sin i$ measurements from the \citet{Glebocki2005} catalog. Using radii from the {\it Kepler} Input Catalog, we converted the starspot periods into equatorial velocities and found a very good agreement between the two samples for spectral types F, G and K. For later type stars (late-K and M dwarfs), the rotational velocities from {\it Kepler} photometry are lower by up to an order of magnitude than implied by the $v \sin i$ catalog, which predominantly contains younger stars in open clusters. For earlier type stars, there is also a disagreement, likely due to incorrect KIC radii.

\begin{figure}[!t]%
\centering
\includegraphics[width=0.9\columnwidth]{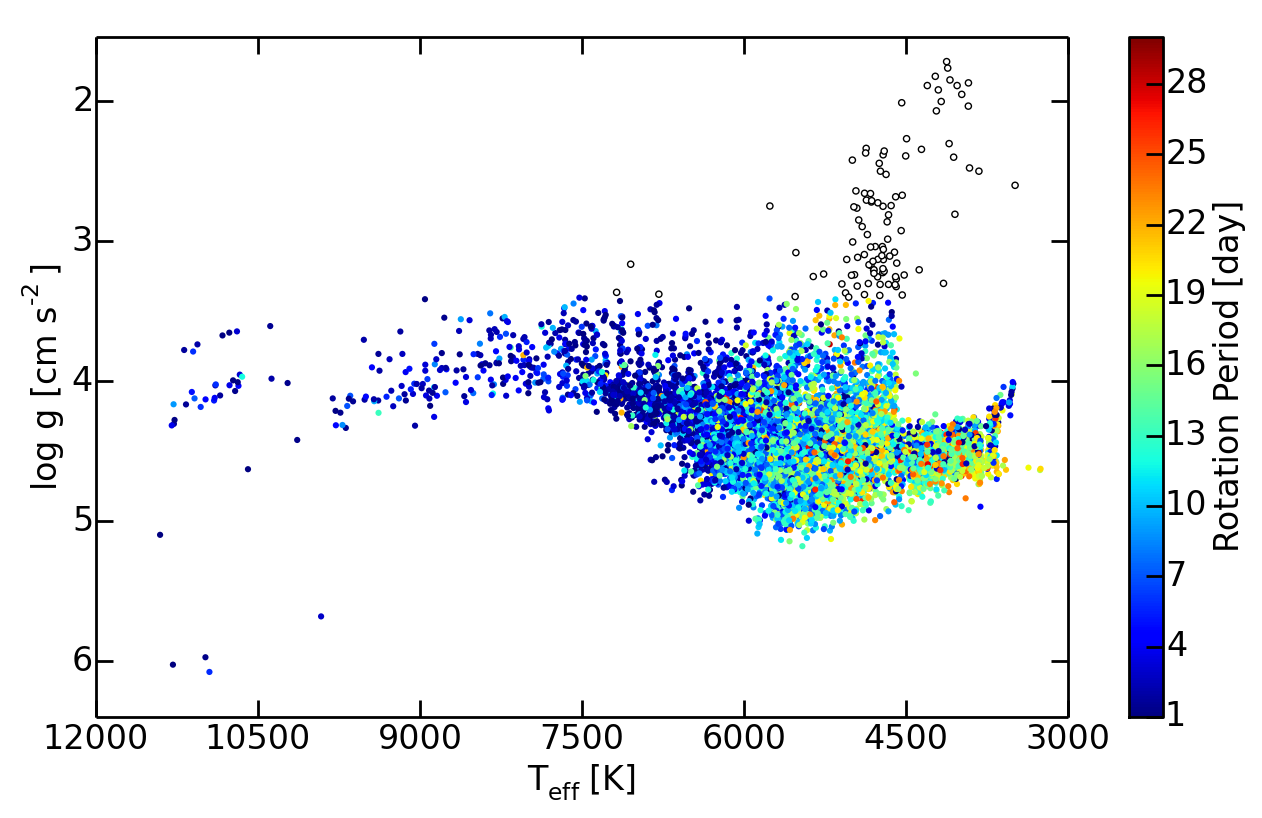}%
\caption{Rotation periods (color scale) versus $\log{g}$ and $T_\mathrm{{eff}}$ read from the {\it Kepler} Input Catalog. The open circles denote stars on the red giant branch with $\log{g} < 3.4$, which are possibly contaminated by p-mode pulsations \citep[see][]{Kjeldsen2011}.}%
\label{fig:rotation_hrd}%
\end{figure}

Figure~\ref{fig:rotation_hrd} shows a ($\log{g}$, $T_{\mathrm{eff}}$)-diagram of the stars for which we measure a rotation period. A strong change in rotation period with temperature along the main sequence is immediately evident around $6000$\,K. This indicates the transition from cool stars with outer convective envelopes to hotter stars with only a shallow convective shell. 

\section{Toward age determination}
\begin{figure}[!t]%
\centering
\includegraphics[width=0.9\columnwidth]{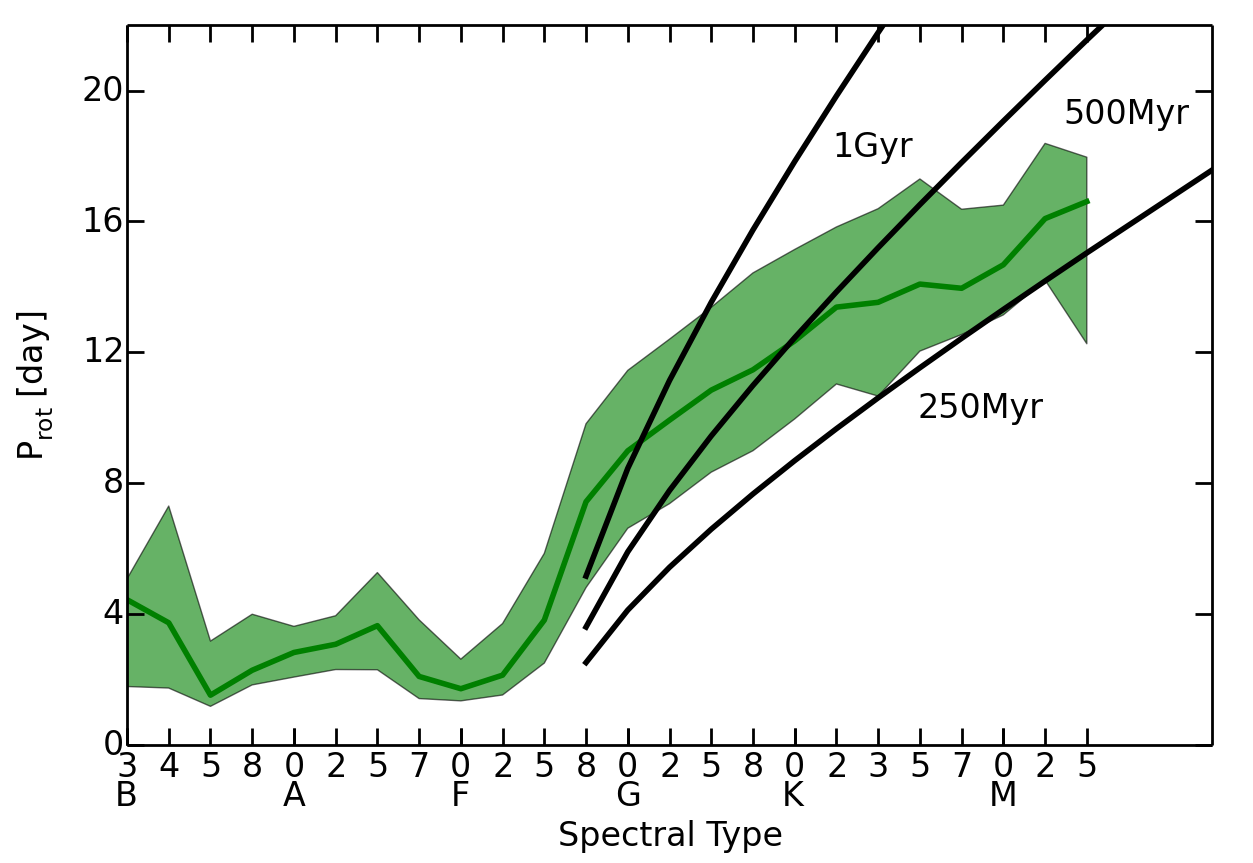}%
\caption{Measured rotation periods as a function of spectral type. The spectral type is derived from the g-r color indexes available from the {\it Kepler} Input Catalog, the shaded area represents the $\pm 34$th percentile values from the median. The isochrones from \citet{Barnes2007} are plotted in solid black for three different ages.}%
\label{fig:p_v_st}%
\end{figure}

Stars rotate slower as they age because of magnetic braking. Cool stars with a deep convective zone exhibit stronger magnetic fields and so lose angular momentum faster via stellar winds. This spin down is described by the relationship between stellar rotation period, stellar age, and B-V color index \citep{Barnes2007}. Three isochrones (250~Myr, 500~Myr, and 1~Gyr) are plotted as a function of rotation period and spectral type (Fig.~\ref{fig:p_v_st}). By comparison with the starspot rotation periods, some information about stellar ages can be obtained. The isochrones indicate that the stars in our sample cover a range of ages up to about 1~Gyr, thus confirming that they are predominantly young active stars.

\acknowledgements M.N., L.G., and H.S. acknowledge research funding by the Deutsche Forschungsgemeinschaft (DFG) under grant SFB 963/1 ``Astrophysical flow instabilities and turbulence'' (Project A18). Funding for the Stellar Astrophysics Centre is provided by The Danish National Research Foundation. The research is supported by the ASTERISK project (ASTERoseismic Investigations with SONG and Kepler) funded by the European Research Council (Grant agreement no.: 267864). This paper includes data collected by the {\it Kepler} mission. Funding for the {\it Kepler} mission is provided by the NASA Science Mission directorate. Some of the data presented in this paper were obtained from the Mikulski Archive for Space Telescopes (MAST). STScI is operated by the Association of Universities for Research in Astronomy, Inc. under NASA contract NAS5-26555. Support for MAST for non-HST data is provided by the NASA Office of Space Science via grant NNX09AF08G and by other grants and contracts.

\bibliographystyle{asp2010}
\bibliography{rotation_paper}

\end{document}